\documentclass[journal=jacsat,manuscript=article]{achemso}

\usepackage{chemformula} 
\usepackage[T1]{fontenc} 

\usepackage{siunitx}
\usepackage{lineno}
\usepackage[colorinlistoftodos]{todonotes}

\newcommand{\F}{Figure }
\author{Kurt H. Tyson}
\affiliation{Department of Physics, Engineering Physics \& Astronomy, Queen's University, Kingston, Ontario, Canada}

\author{James R. Godfrey}
\affiliation{Department of Physics, Engineering Physics \& Astronomy, Queen's University, Kingston, Ontario, Canada}

\author{James M. Fraser}
\email{jf9@queensu.ca}
\affiliation{Department of Physics, Engineering Physics \& Astronomy, Queen's University, Kingston, Ontario, Canada}

\author{Robert G. Knobel}
\email{knobel@queensu.ca}
\affiliation{Department of Physics, Engineering Physics \& Astronomy, Queen's University, Kingston, Ontario, Canada}

\title[An \textsf{achemso} demo]
  {Localized Gradual Photomediated Brightness and Lifetime Increase of Superacid Treated Monolayer \ch{MoS2}}

\abbreviations{S-TMDs, TFSI, PL, PLQY, 2D, SPCL, CW, AFM, ROI, IRF, IPA, DI, PPC, PDMS, DCE, DCB, CCD, OPCPA, FWHM,}
\keywords{transition-metal dichalcogenides, lifetime dynamics, superacid treatment, photoluminescence, photo-enhancement}

\begin{document}

\begin{tocentry}
    \includegraphics[width=9cm, height=3.5cm]{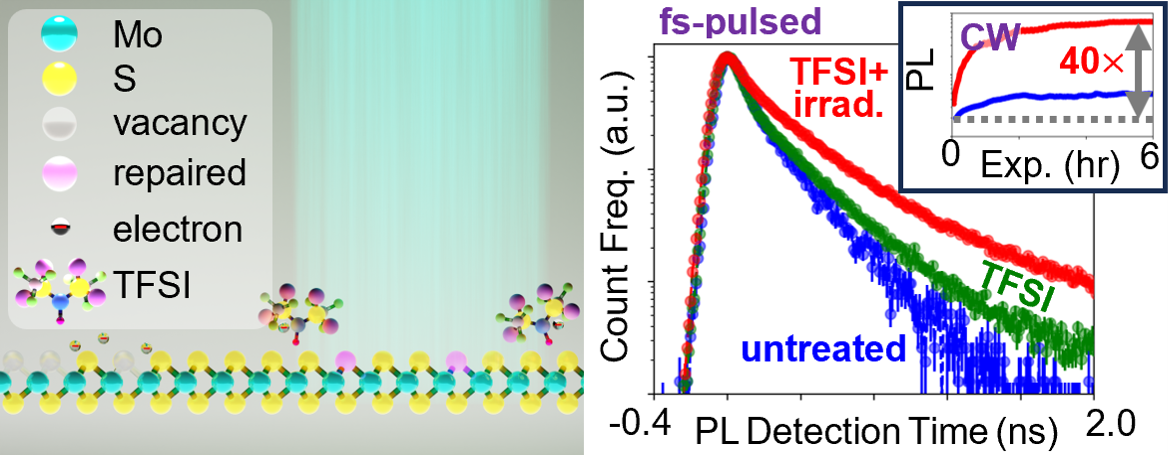}

\end{tocentry}

\begin{abstract}
   Monolayer semiconducting transition metal dichalcogenides (S-TMDs) have been extensively studied as materials for next-generation optoelectronic devices due to their direct band gap and high exciton binding energy at room temperature. Under a superacid treatment of bis(trifluoromethane)sulfonimide (TFSI), sulfur-based TMDs such as \ch{MoS2} can emit strong photoluminescence (PL) with photoluminescence quantum yield (PLQY) approaching unity. However, the magnitude of PL enhancement varies by more than two orders of magnitude in published reports. A major culprit behind the discrepancy is sulfur-based TMD's sensitivity to above band-gap photostimulation. Here, we present a detailed study of how TFSI-treated \ch{MoS2} reacts to photostimulation with increasing PL occurring hours after constant or pulsed laser exposure. The PL of TFSI-treated \ch{MoS2} is enhanced up to 74 times its initial intensity after 5 hours of continuous exposure to 532\,$\unit{\nano m}$ laser light. Photostimulation also enhances the PL of untreated \ch{MoS2} but with a much smaller enhancement. Caution should be taken when probing \ch{MoS2} PL spectra as above-bandgap light can alter the resulting intensity and peak wavelength of the emission over time. The presence of air is verified to play a key role in the photostimulated enhancement effect. Additionally, the rise of PL intensity is mirrored by an increase in measured carrier lifetime of up to $\sim$400\,$\unit{\pico s}$ consistent with the suppression of non-radiative pathways. This work demonstrates why variations in PL intensity are observed across samples and provides an understanding of the changes in carrier lifetimes to better engineer next-generation optoelectronic devices. 
\end{abstract}


\section{Introduction}

Two-dimensional (2D) semiconducting materials such as transition-metal dichalcogenides (TMDs) hold promise for ultrathin and flexible optoelectronic devices due to their single molecular layer thinness and lack of out-of-plane dangling bonds \cite{Mak2010, Siao2018, Laturia2018}. More importantly, at the monolayer limit, semiconducting TMDs (S-TMDs) such as \ch{MoS2} exhibit a direct band gap at the K-K’ location of the Brillouin zone \cite{Mak2010} with large exciton binding energy (several hundred meV) stable at room temperature \cite{Hill2015}. In combination with their ability to form heterostructures without the need of lattice matching \cite{Li2018, Lin2014}, many atomically thin optoelectronic devices such as photodetectors \cite{Lopez-Sanchez2013, Liu2021}, light emitters \cite{Kwon2019, Hotger2021, Lien2018}, lasers \cite{Salehzadeh2015}, and photovoltaics \cite{Singh2017, Tsai2014, Wi2014} have been demonstrated.

Two key challenges hindering S-TMD based devices are their inherently high defect density and susceptibility to their environment \cite{Aryeetey2020, Mirabelli2016, Park2018}. This causes the untreated photoluminescence quantum yield (PLQY) to be low (0.01-6\%) \cite{Amani2015a, Mak2010} depending on the quality of the prepared material. Reduction of PLQY greatly limits efficient light emission and absorption. In \ch{MoS2}, sulfur vacancies are the dominant defect observed in both chemical vapor deposition and mechanically exfoliated crystals \cite{Roy2018, Hong2015, Pain2022}. These defects create midgap states which lead to non-radiative recombination of neutral and charged excitons \cite{Bretscher2021, Lu2018, Goodman2017}, reducing the PLQY. Attempts to repair the effects of these defect sites have been reported including passivating agents \cite{Bretscher2021}, chemical dopants \cite{Tanoh2019, Mouri2013, Li2021}, thermal annealing \cite{Su2017}, and electrostatic gating \cite{Lien2019}. One of the most successful methods involves a superacid, bis(trifluoromethane)sulfonamide (TFSI) treatment which demonstrated increased PLQY to near unity at low fluences\cite{Amani2015a, Amani2016}. Although the PL enhancement is thought to involve passivation and p-doping, the exact mechanism is heavily debated \cite{Bretscher2021, Amani2015a, Goodman2017, Li2021, Tanoh2021}. Unfortunately, the PL enhancement using TFSI acid is highly variable between studies and samples, ranging from 10$\times$ to over 200$\times$ PL \cite{Amani2016, Yamada2020, Tanoh2019, Cadiz2016a, Alharbi2017, Roy2018}. In some cases, the difference is as stark as two orders of magnitude despite using a similar methodology. This raises questions of what factors can cause such discrepancy. Recently, interest has been drawn towards \ch{MoS2}'s photosensitivity where ultraviolet light and hydrogen annealing with light-based "treatments" have been shown to greatly increase PL in ambient conditions \cite{Yamada2020, Sivaram2019, Ardekani2019, Yamada2021}. In this report, we show that the light used for PL metrology can in fact lead to the photo-enhancement over a period of hours, a possible reason for the large discrepancies in PL intensity in previous work. In addition, we observe not only greater PL enhancement with increasing excitation power but also faster rates of growth, but the enhancement requires air to be present, as verified by our PL measurements in vacuum. 

To further probe the origin of the photostimulation, we perform ultrafast single-photon counting lifetime (SPCL) measurements. Increased PL could be due to a reduction in non-radiative pathways, an increase in the number of injected excitons, or a combination of both. In TFSI-treated samples, we observe the PL lifetime increases with photostimulaton, consistent with reduced non-radiative recombination. We observe this improving PL lifetime in a high exciton density regime, which is of particular interest for high-current device applications.

\section{Results and discussion}

\subsection{Localization and Enhancement of Steady-State Photoluminescence}

\begin{figure}[ht!]
\centering\includegraphics[width = 14cm, trim={0 0.45cm 0 0.2cm},clip,keepaspectratio=true]{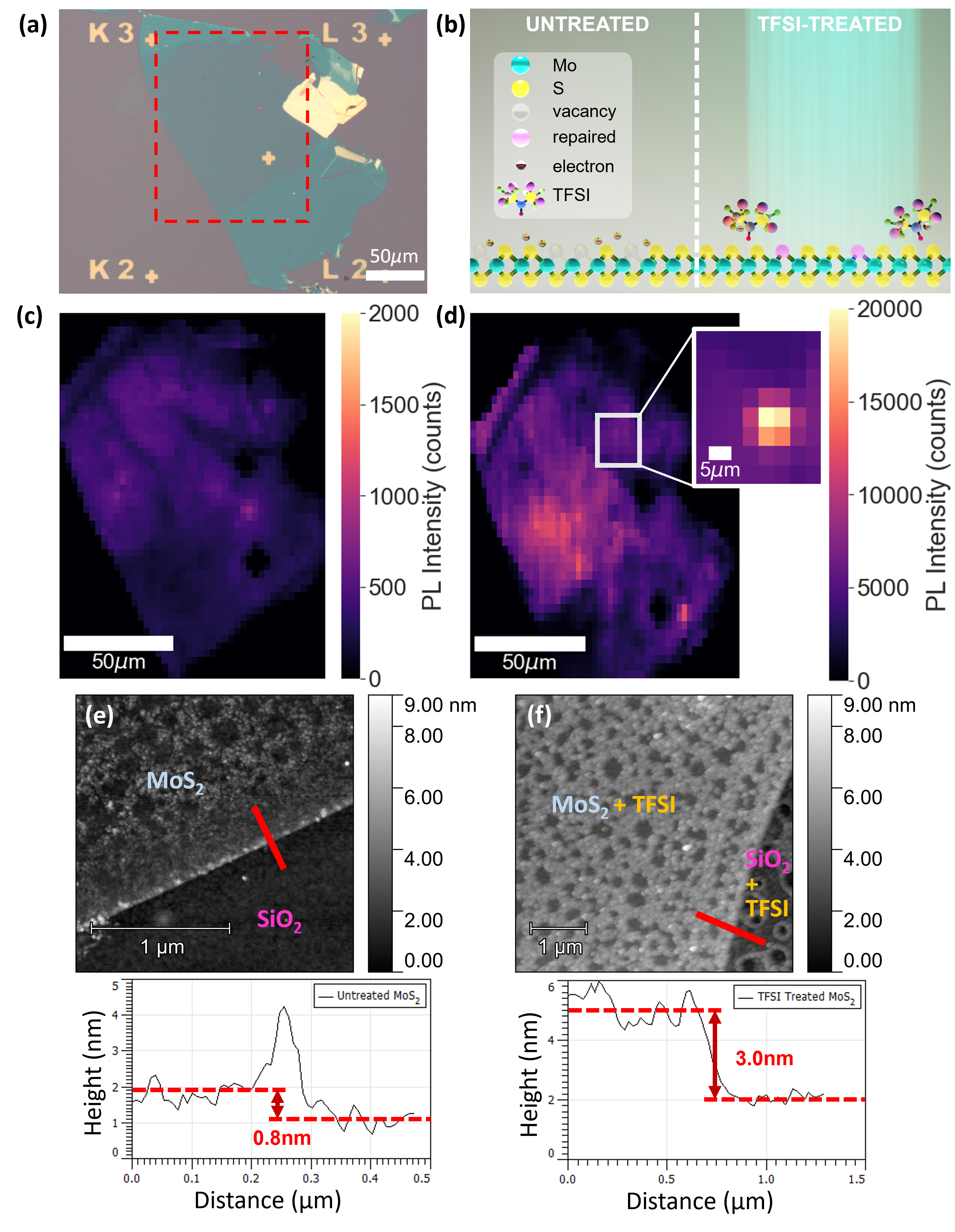}
\caption{ (a) Optical microscopy image of monolayer \ch{MoS2} flake. (b) A schematic of TFSI-\ch{MoS2} system. The untreated side shows an abundance of vacancies and excess electrons causing n-doping, while the TFSI-treatment with laser exposure is shown to facilitate defect substitution and charge neutralization. PL map of the (c) untreated flake and (d) TFSI-treated flake shows on average greater than $10\times$ PL enhancement with minimal light exposure. Inset is the 1\,hr exposure of $532\,\unit{\nano m}$ laser light ($25\,\unit{\micro W}$) on TFSI-treated \ch{MoS2} further enhancing the PL intensity locally. AFM images and sample height cross-sections (corresponding to red line shown in image) before (e) and after (f) TFSI-treatment.}
\label{fig:SampImage_PLmap_AFM}
\end{figure}

We first compare changes in PL between TFSI-treated and untreated exfoliated \ch{MoS2} subjected to continuous laser exposure. A `gold-assisted' \cite{Desai2016} exfoliation method was used to obtain large area (>$100\,\unit{\micro m}$ side length) monolayer \ch{MoS2} flakes (\F\ref{fig:SampImage_PLmap_AFM}a), allowing many consecutive PL measurements to be conducted on the same sample. This method takes advantage of the strong surface binding energy of \ch{MoS2} to gold \cite{Velicky2020} in order to mechanically cleave monolayer \ch{MoS2} flakes onto a Si/Si\ch{O2} substrate. Immediately after sample preparation, a $532\,\unit{\nano m}$ continuous-wave (CW) laser (4\,\unit{\micro m} diameter spot size) is used to measure PL in ambient air. 

 A 2D image of integrated PL intensity was created by taking a raster scan over the flake's surface. The PL intensity and peak wavelength (\F S1) were relatively uniform over the area of the flake (\F\ref{fig:SampImage_PLmap_AFM}c). Post TFSI-acid, the PL intensity increased ($\sim$10$\times$ enhancement) as expected (\F\ref{fig:SampImage_PLmap_AFM}d). Surprisingly, continuous exposure to $532\,\unit{\nano m}$ light further increased PL intensity (up to an additional 10$\times$). The enhancement was localized to the spot size of the laser and occurred in both untreated and TFSI-treated \ch{MoS2}. The enhancement of PL did not appear to diffuse or affect the surrounding areas of the sample, staying localized at the focus of the laser. Without constant exposure, the laser-mediated PL enhancement effect fades within a few hours while the TFSI enhancement fades over a few days in air.

Atomic force microscopy (AFM) measurements were taken before and after TFSI treatment (\F\ref{fig:SampImage_PLmap_AFM}e,f). The height profile of an untreated monolayer \ch{MoS2} flake is $0.8\,\unit{\nano m}$ which confirmed exfoliation of a single molecular layer was achieved \cite{Lauritsen2007}. All PL measurements were taken near the center of the \ch{MoS2} samples to minimize edge effects due to strain, tearing, or residual build-up. After TFSI treatment, AFM shows the TFSI preferentially deposits on \ch{MoS2}, with a $3.0\,\unit{\nano m}$ height increase relative to the acid layer on the adjacent substrate post-annealing. As the TFSI solution dries, it concentrates as a thin film over the monolayer, preserving the elevated PL for a few days. The TFSI coating has a few nanometer height variations between the rough features seen over the monolayer surface. The uneven coating of the TFSI may partially explain why variations in PL intensity and center wavelength are measured. 

\begin{figure}[ht!]
\centering\includegraphics[width = 16.5cm, trim={0.75cm 0 0 0},clip,keepaspectratio=true]{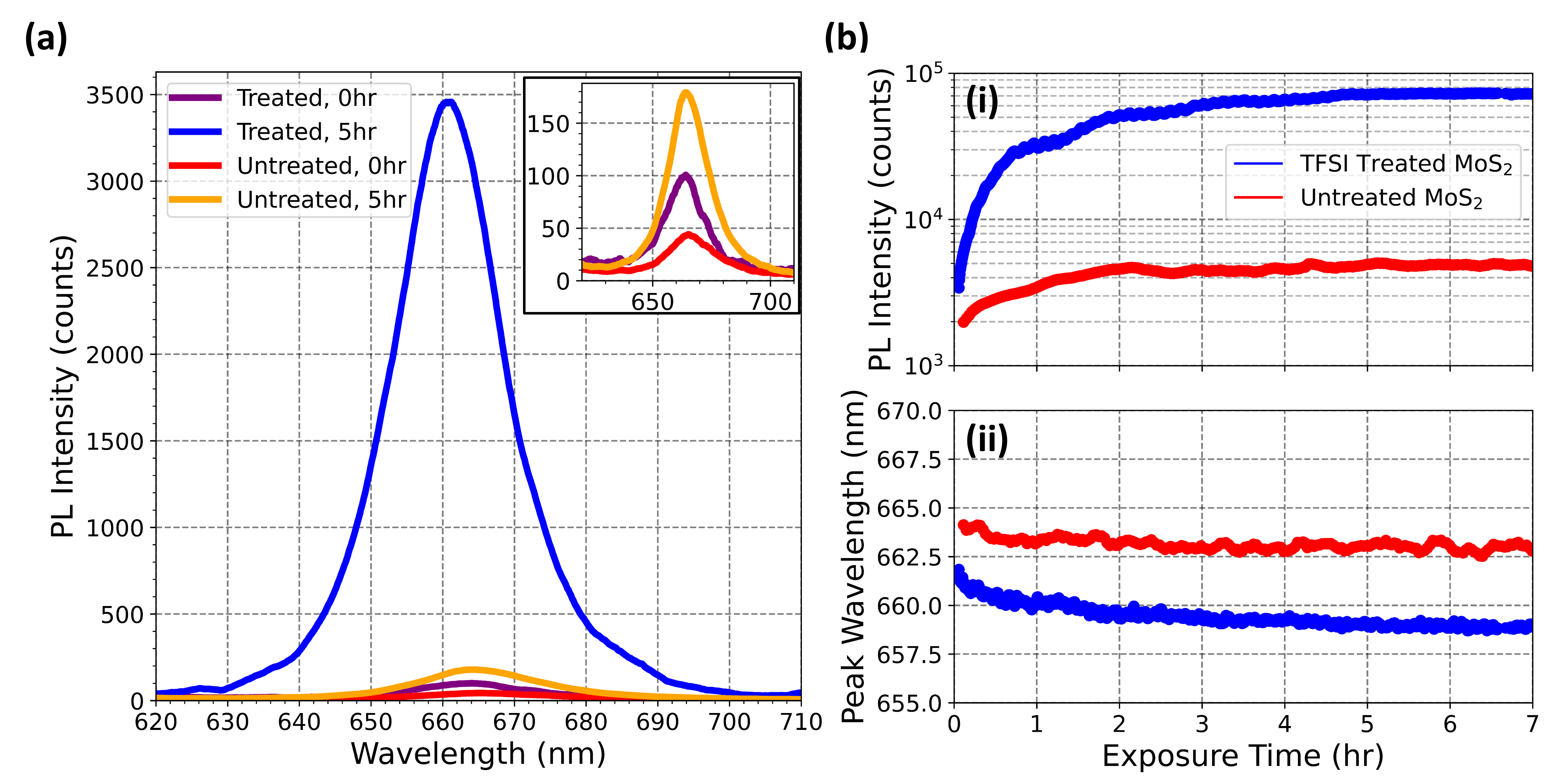}
\caption{(a) PL spectra of untreated and TFSI-treated \ch{MoS2} at 0\,hr and 5\,hr of exposure to a 532\,nm CW laser. Inset: PL of the untreated sample and unexposed TFSI-treated sample magnified around the main spectral feature. (b) Integrated PL intensity (i) and peak wavelength of PL spectra (ii) of untreated and TFSI-treated \ch{MoS2}.}
\label{fig:PLspectra}
\end{figure}

 To characterize the gradual PL enhancement with low-power CW light, we tracked its evolution with time and intensity. \F\ref{fig:PLspectra}a shows the PL spectra before and after treatment ($25\,\unit{\micro W}$ laser power, \qty{532}{nm}, 4\,\unit{\micro m} diameter spot size). After locating a region of interest (ROI), successive PL spectra measurements were recorded with the ROI exposed constantly to the CW laser. The PL continued to increase hours after first exposure, lasting in some cases over 20\,hours before decreasing. The total magnitude of enhancement is consistent with other reported TFSI-treated \ch{MoS2} enhancement values \cite{Yamada2021a, Roy2018}. The greatest rate of increase in all cases occurred at initial exposure, with the intensity typically plateauing hours later. Comparing unexposed ($0\,\unit{hr}$) untreated \ch{MoS2} PL with $7\unit{hr}$ laser-irradiated and TFSI-treated \ch{MoS2} combined, a total enhancement factor of $\sim$74$\times$ is observed. A typical increase in PL intensity ranges from 3-15$\times$ with only TFSI treatment. Photo-enhancement of TFSI-treated \ch{MoS2} further increases PL up to 10$\times$ of the TFSI-treated PL intensity. A modest yet significant 2-3$\times$ photo-enhancement occurs in non-TFSI treated \ch{MoS2}. 

The PL spectra show a peak in untreated \ch{MoS2} at 664.5\,$\unit{\nano m}$ (1.866\,eV) typically attributed to neutral exciton recombination \cite{Christopher2017a} which blue shifts to 662.8\,$\unit{\nano m}$ (1.871\,eV) after 7\,hr of CW light exposure (\F\ref{fig:PLspectra}b(ii)). Similarly, in TFSI-treated \ch{MoS2} a blue shift from 662.3\,$\unit{\nano m}$ (1.872\,eV) to 658.5\,$\unit{\nano m}$ (1.883\,eV) is observed. A shoulder on the red side of the PL spectra can be seen in untreated \ch{MoS2} due to trion (charged excitons; A$^-$) emissions \cite{Christopher2017a}. When treated by light or by TFSI-acid, the exciton contribution to the overall PL further dominates compared to trions. The result is the expected blue shift and narrowing of the PL spectra. This can be caused by oxidation or p-doping of \ch{MoS2} where naturally occurring sulfur vacancies are passivated \cite{Ardekani2019}. Reducing the negative carrier concentration in naturally n-doped \ch{MoS2} would make the formation of trions less likely, leading to greater neutral exciton formation and increased radiative recombination \cite{Amani2015a}. A reduction in trions and non-radiative pathways, as well as an increasing neutral exciton population, would explain the increase in PL intensity and blue shift during TFSI treatment and photo-enhancement. Yamada \textit{et al.} report a similar photoactivated enhancement  where TFSI-treated \ch{MoS2} PL was shown to increase after wide-area UV light exposure but unchanged for untreated \ch{MoS2} \cite{Yamada2020}; Yamada and coauthors attributed this photo-mediated increase in PL to TFSI's chemical effects. In contrast to their results, we also observe a PL enhancement in untreated \ch{MoS2} over hours. Thus photo-enhancement is not limited to TFSI treatment (\F\ref{fig:PLspectra}b), though the underlying mechanisms may be different.

The exposure-dependent response changing the PL by over an order of magnitude suggests that caution should be taken when quantifying PL enhancement factors or PLQY, or otherwise in applications that rely on consistent absorption or emission properties (i.e., detectors and light sources, respectively). As PL studies of \ch{MoS2} typically use above-bandgap CW excitation, our observations suggest it is important to track the site-specific laser exposure or segment the measurements by time since this greatly impacts the resulting PL intensity observed. Studies that quantify enhancement factors due to external treatments (such as acid treatments, gating, plasmonic nanoparticles, etc) need to be especially careful to control for localized laser irradiation.

\subsection{Power Dependence of Laser-Mediated Enhancement}

\begin{figure}[ht!]
\centering\includegraphics[width = 16.5cm, trim={0.5cm 0.5cm 0 0},clip,keepaspectratio=true]{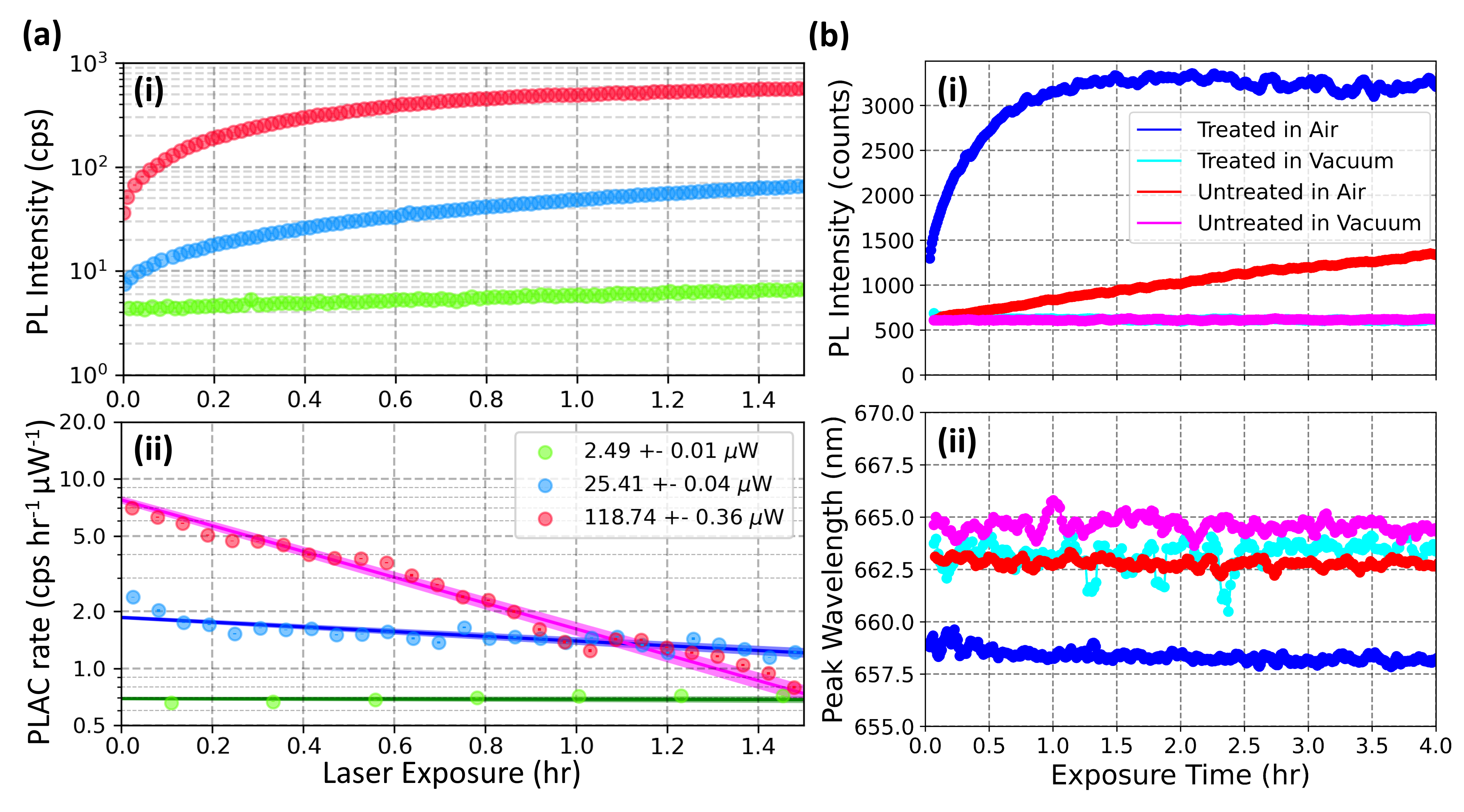}
\caption{(a-i) PL from TFSI-treated \ch{MoS2} as a function of 532\,\unit{\nano m} laser exposure time at 118.7\,$\unit{\micro W}$ (red), 25.4\,$\unit{\micro W}$ (blue), and 2.5\,$\unit{\micro W}$ (green). (a-ii) PLAC growth rates calculated from (a-i), regression fit lines shown with 3$\sigma$ error shading. The initial rate of PLAC increase is greater with higher power; however the rate of increase also quickly drops at higher power. (b-i) PL intensity and (b-ii) peak wavelength of untreated and TFSI-treated \ch{MoS2} under 25\,$\unit{\micro W}$ of CW 532\,\unit{\nano m} excitation, over time in air and in vacuum. No PL enhancement or spectral shift occur in vacuum.}
\label{fig:PowerDep_Rates_Vacuum}
\end{figure}

PL at various excitation powers (\F\ref{fig:PowerDep_Rates_Vacuum}a(i)) is collected at multiple locations on a large TFSI-treated monolayer \ch{MoS2} flake ($\sim$$200\,\unit{\micro m}$\,$\times$\,$200\,\unit{\micro m}$). All measurements were taken within $48\,\unit{hr}$ of TFSI treatment, as such treatment has been shown to last a few days in ambient air before slowly degrading \cite{Yamada2021a}. PL action cross-section growth rates (henceforth PLAC rates) are calculated as the average rate of change in PL intensity obtained from linear regression on each 1\,min time window in \F\ref{fig:PowerDep_Rates_Vacuum}a(i), also show a dependence on incident power (\F\ref{fig:PowerDep_Rates_Vacuum}a(ii)). As this analysis requires numerical differentiation, PL intensity data is first lowpass filtered to reduce noise (LOESS filter).

At the highest power ($118.7\,\unit{\micro W}$), the PLAC rate dropped to $\sim$50\% of its initial value within $30\,\unit{min}$, while at intermediate ($25\,\unit{\micro W}$) and  low ($2.5\,\unit{\micro W}$) powers, the growth rate is $\sim$50\% of its initial value at $1.5\,\unit{hr}$ and $7\,\unit{hr}$ respectively. Higher power exposure provides a greater number of above bandgap photons to assist in photomediated processes (e.g., OH radical dissociation from \ch{H2O}, sulfur and oxygen atoms filling in vacancies). As sulfur vacancies are repaired and midgap states are removed, the initial increase in PL intensity is rapid. As more of these defect sites are repaired, there become fewer sites available to further increase PL. 

It should be noted that PL as reported here depends on both incident flux as well as the repair process that may also depend on incident flux; the PLAC rates aid in detangling these two effects. At t=0, higher fluence corresponds to higher rates, consistent with the model that repair relies on incident flux (\F\ref{fig:PowerDep_Rates_Vacuum}a(ii)). Interestingly, the PLAC growth rate appears to decrease quasi-exponentially with time (straight line on semi-log plot shown in \F\ref{fig:PowerDep_Rates_Vacuum}a(ii)) suggesting that the higher fluence is indeed depleting the number of defects available for repair at a faster rate. We consider a simple model where a finite number of defects are available for repair, and the repair probability depends on incident fluence. Such a model predicts that the exponential decay of PLAC rate scales linearly with fluence (supplemental information). Linear regression fits to the moderate and high power agree with this model to within 15\%. We note that a larger survey of different locations on each sample would be required to obtain a quantitative analysis of the true power dependence of this effect (to overcome the issue of sample non-uniformity).

\subsection{Effects of Ambient Air on Photo-Enhancement}
Previous reports show that the presence of air plays a role in the PL enhancement process \cite{Bera2018, Yamada2020, Sivaram2019}. Untreated and TFSI-treated \ch{MoS2} samples are first irradiated with CW $532\,\unit{\nano m}$ laser light over several hours within a vacuum chamber containing ambient air. PL enhancement rate for the TFSI-treated \ch{MoS2} increases with laser exposure power (\F\ref{fig:PowerDep_Rates_Vacuum}a, discussed below). The vacuum chamber was then pumped down to $<$$10^{-4}\,\unit{Torr}$, and previously unexposed areas on the same monolayer flake were exposed to light. Both untreated and TFSI-treated \ch{MoS2} in vacuum show no change in integrated PL or peak wavelength over $4\,\unit{hr}$ of light exposure (25.4\,\unit{\micro W}), in contrast to in-air behaviour (\F\ref{fig:PowerDep_Rates_Vacuum}b). These results confirm that air plays an important role in both TFSI-treated enhancement and photo-enhancement. Samples treated by TFSI days in advance and stored in ambient still display rapid PL growth from light exposure starting from a similar baseline PL, meaning that brightening is not dependent on the amount of time exposed to ambient gases (e.g., \ch{H2O}, \ch{O2}) but is instead photomediated. We also extract growth rates from \F 3b(i), to compare between acid-treated and untreated samples in air under equivalent excitation conditions and measurement parameters. Comparing the initial growth rate for the acid-treated sample (up to 10\,min) to the average growth rate of the untreated over the entire 4\,hr, the growth rate is 24$\times$ greater.

More than one mechanism may be contributing to the enhancements observed from TFSI, photoirradiation, and ambient air. The most common defect in \ch{MoS2} are single sulfur vacancies which introduce mid-gap states that reduce overall radiative emissions \cite{Roy2018, Bretscher2021}. Firstly, photoirradiation of \ch{MoS2} in air has been reported to promote chemisorption and physisorption of \ch{O2}, \ch{H2O}, and OH radicals which passivate sulfur vacancies and act as a p-dopant \cite{Birmingham2018}. Highly bound excitons interact with physisorbed \ch{H2O} molecules via empty antibonding orbital (\ch{H2O}$^*$) in order to facilitate the dissociation into OH \cite{Sivaram2019}. This is supported by the observed blue-shifting and increase in PL intensity in our untreated \ch{MoS2} photo-enhancement. Secondly, TFSI compared to other acids (e.g., Oleic acids, \ch{H2SO4}, etc) \cite{Tanoh2019, Wang2022, Kiriya2018} is proven to be more effective at increasing PL intensity due to its ability as a p-dopant to suppress non-radiative trions, but also because it can directly repair sulfur vacancies with sulfur atoms \cite{Zeng2018}. Roy et al. comment that TFSI dissociates into \ch{SO2} and \ch{CF3}-\ch{N}-\ch{SO2}-\ch{CF3} and finally into an absorbed S atom (and \ch{O2} molecule) which can repair vacancies\cite{Roy2018}. However, this requires an activation energy of 0.55\,eV to occur. The needed activation energy could come from photoirradiation. Note though that we observe that removing air from the TFSI-treated \ch{MoS2} returns PL to the same level as untreated (\F\ref{fig:PowerDep_Rates_Vacuum}b)) showing the repair process is not permanent (i.e., more likely physisorbed than chemisorbed). Other possible pathways related to the presence of air do exist: photomediated OH radicals have been thought to alter the hydrophilicity of the \ch{MoS2} surface allowing for increased interaction by the hydrophilic TFSI molecules\cite{Yamada2021}. \ch{H2O} and \ch{O2} by themselves could cause some PL enhancement in untreated \ch{MoS2} through physisorption or chemisorption to sulfur vacancies.  

\subsection{TFSI-Treated MoS2 Shows Improved PL Lifetime In High Exciton Density Regime}
In order to better understand the origin of the enhanced PL, we perform time-correlated single photon counting lifetime (SPCL) measurements. Here it is the short pulse $550\,\unit{\nano m}$ (necessary for lifetime measurements) that serves as the light source for photoactivation. The incident flux ($13.4\,\unit{\nano W}$) needed to be highly constrained to avoid sample damage from extreme instantaneous irradiance. Even with such low fluences, PL enhancement of $\sim$1.5-4$\times$ for treated samples (\F S5 and \F S6) and $\sim$1.5-2$\times$ for untreated was observed after 7 hours of exposure. 
Exposure times sufficiently long to capture full enhancement were not possible; a naive model assuming defect repair rate scales with total incident fluence predicts that 13.4\,\unit{\nano W} exposure would require laser exposure for longer than a month before fully enhanced PL was realized. This is not surprising since CW-exposure with an incident flux 200$\times$ higher (2.5\,$\unit{\micro W}$) was still slightly increasing after 7 hours.  Note also that the PL enhancement still benefitted from TFSI treatment: TFSI-treated samples had approximately an order of magnitude higher PL intensity than the untreated samples. 

\begin{figure}[ht!]
\centering\includegraphics[width = 16.5cm, trim={0.35cm 0 0 0},clip,keepaspectratio=true]{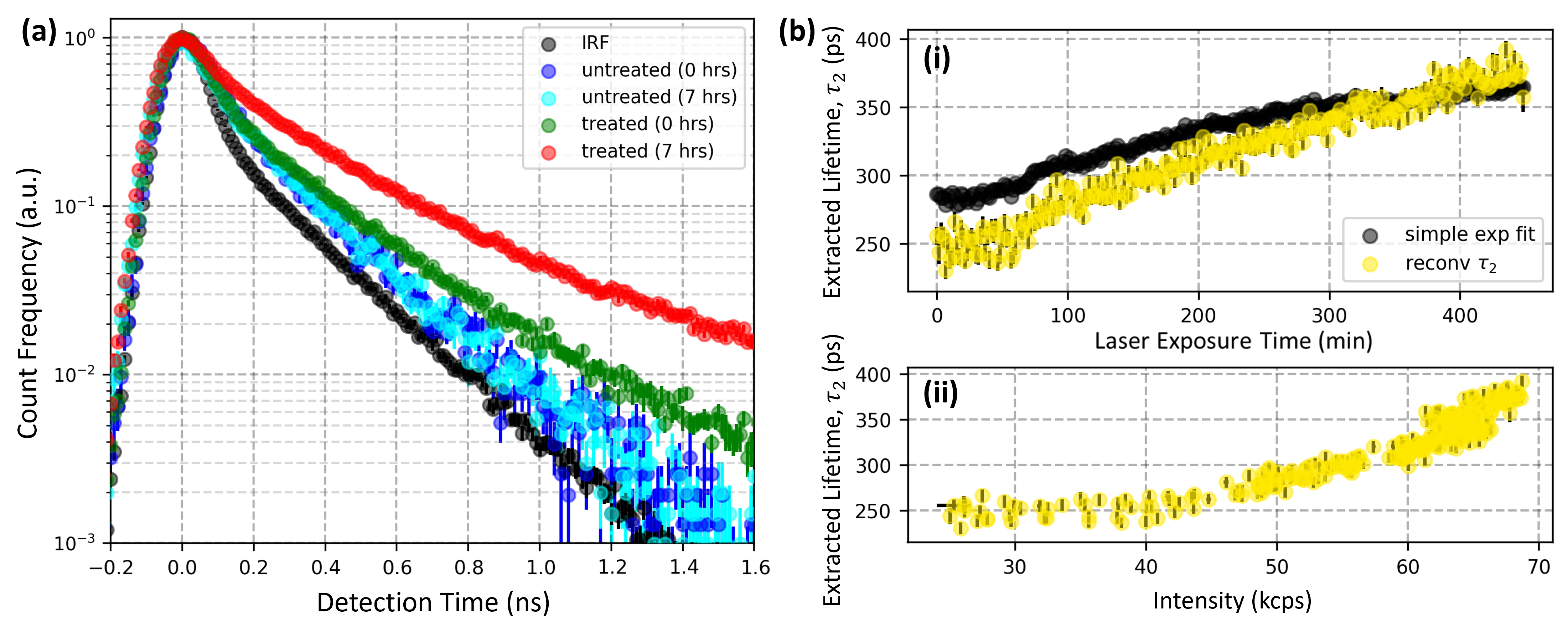}
\caption{(a) SPCL histograms for untreated and TFSI-treated \ch{MoS2} lifetime at 0\,hr and 7\,hr of 550\,nm pulsed laser exposure (13.4\,nW average power, 1\,MHz rep rate, 300\,fs pulse duration). (b-i) Extracted photoluminescence lifetime via exponential fit and reconvolution vs. pulsed laser exposure time. (b-ii) Extracted lifetime vs. integrated intensity under pulsed excitation.}
\label{fig:Ultrafast}
\end{figure}

Even with this modest PL enhancement, PL lifetime is observed to increase with increasing laser exposure in TFSI-treated samples (\F\ref{fig:Ultrafast}). The untreated samples showed no detectable change in the measured carrier lifetimes with $7\,\unit{hr}$ of continued laser exposure but TFSI treatment immediately shows increased lifetime (\F\ref{fig:Ultrafast}a). Values are extracted from the SPCL histograms using two fitting methods: (1) a standard single exponential fit to the region after the main peak of the instrument response in the histogram (between $150\,\unit{\pico s}$ and 1.5\,\unit{\nano s}), and (2) a reconvolution method wherein a biexponential function is fit to the histogram data via regression, after being discretized and convolved with the instrument response function (IRF) \cite{Smith2017}. From the simple exponential fit, we obtain lifetimes of the untreated samples of $\sim$225\,\unit{\pico s}, very close to the lifetime obtained by direct exponential fitting of the IRF. Shown in \F\ref{fig:Ultrafast}b, the simple exponential fit recovers a lifetime ranging between 280-360\,\unit{\pico s} for the TFSI-treated sample, though the extent to which the IRF is influencing the measurement is unclear.

To remove the influence of the IRF, we employ the biexponential reconvolution method. We use a fit function treating the system as having two independent levels with competing exponential decays:

\begin{equation}
    I(\tau_1, \tau_2, t_0, A_1, A_2) = A_1\exp{ \biggl( \frac{-(t-t_0)}{\tau_1} } \biggr) +  A_2\exp{ \biggl( \frac{-(t-t_0)}{\tau_2} } \biggr) \label{eqn:2LT}
\end{equation}
where $\tau_i$ and $A_i$ are the lifetimes and amplitudes, respectively, of the two competing exponential decays, and $t_0$ is the time of the initial excitation of the system. Eq.\ref{eqn:2LT} is combined with our experimentally acquired IRF via convolution, and the fitting parameters are optimized to minimize the mean-squared error, allowing us to extract experimental PL lifetimes with minimal influence of the IRF.

The biexponential reconvolution fit has a fast component $\tau_1$ below 100\,$\unit{\pico s}$, as such fast lifetimes have been reported in monolayer \ch{MoS2}\cite{Moody2016}, while the slower component $\tau_2$ is allowed to >100\unit{\,\pico s}. This approach has the advantage over other convolution fitting procedures (e.g., fitting a convolution of a Gaussian with an exponential \cite{Hong2020}) in that it allows for an arbitrarily shaped detector response to be removed from the data. The reconvolution method for the untreated samples required the long lifetime component disabled to avoid over-fitting,  recovering a single lifetime from untreated samples of between 50-80\,$\unit{\pico s}$. As typical reported lifetimes for untreated \ch{MoS2} range from 0.3-50\,$\unit{\pico s}$ \cite{Moody2016, Amani2015a}, this suggests for this particular data set (which is considerably noisier than results for TFSI-treated samples) the lower measurement limit of this approach is $\sim$80\unit{\pico s}. \F\ref{fig:Ultrafast}b shows time-resolved SPCL data for one location on a treated sample which had an extracted $\tau_2$ lifetime ranging from 250-400$\unit{\pico s}$, with the observed lifetime increasing continuously over more than $7\unit{hr}$ of exposure to the pulsed excitation. Other locations on the sample allowed extraction of a $\tau_2$ lifetime typically ranging from 200-400$\unit{\pico s}$ exhibiting similar lifetime increases with continued laser exposure (\F S7). 

\F\ref{fig:Ultrafast}b(ii) shows that this increase in lifetime also accompanies an increase in the PL intensity, suggesting that the mechanism by which the brightness is increasing subdues faster non-radiative pathways, and is not solely due to an increase in absorption. Observation of the lifetime changing commensurately with the intensity (shown explicitly in \F S5) is consistent with an improving quantum yield, which seems to be through the dynamic light-mediated process we observed in acid-treated samples. Although PLQY and exciton lifetimes are greatest at low fluence, it is worth noting that the longer $\sim$$ 400\,\unit{\pico s}$ lifetimes are on par with other literature results when considering high pump fluences \cite{Yamada2020}. Being in a regime of high exciton density (590\,\unit{\micro m^{-2}}; based on an absorption of $\sim$5\% at 550\,$\unit{\nano m}$ for TFSI-treated \ch{MoS2}\cite{Goodman2020} and measured energy density of 425\,\unit{ \nano J /\centi m^{2}}) is a regime of interest for \ch{MoS2}--based light sources and detectors as such a regime is typically where a device is being pushed to its operational limit. 

It is understood that a possible mechanism for the photo-enhancement of untreated MoS$_2$ is that light exposure mediates the chemisorption and/or physisorption of oxygen from water molecules to sulfur vacancies through exciton generation \cite{Sivaram2019}. Repairing sulfur vacancies through oxygen substitution would remove dark midgap states that are deleterious to the desired PL response \cite{Bretscher2021}. This is possible even without the presence of TFSI acid which explains our observed mild enhancement in our untreated samples. In the treated samples, there may be a second enhancement mechanism acting in concert with the water adsorption mechanism: TFSI can also dissociate into sulfur atoms in the presence of water\cite{Roy2018}, which can further repair vacancy sites in \ch{MoS2}. It appears that either TFSI breakdown or the water adsorption mechanisms are greatly catalyzed by the excitation light, or the water adsorption mechanism is further mediated by the TFSI in the presence of excitation light, though we cannot distinguish between these mechanisms. In our TFSI-treated samples, the observation of increased \ch{MoS2} carrier lifetime and PL combined with the requirement of ambient air containing water vapour further supports the hypothesis of photo-catalyzed defect repair. The hypothesis of only the water adsorption mechanism occurring in untreated samples while multiple mechanisms occur in the treated samples is supported by the observed initial PL growth rate being 24$\times$ higher in TFSI-treated samples at equivalent fluence (\F 3b(i)).

\section{Conclusion}
We have shown enhancement of PL intensity and increased lifetime in TFSI-treated \ch{MoS2} over hours of above-bandgap laser exposure. This process is observed in both superacid-treated and untreated \ch{MoS2}, though modest in untreated samples, indicating the mechanisms at play are not limited to acid treatments alone. The considerably stronger photo-enhancement effect in TFSI-treated \ch{MoS2} may be due to a combination of mechanisms to repair sulfur defects and p-dope the monolayer. We have also observed real-time changes to the lifetime dynamics of treated \ch{MoS2} as PL is enhanced. Photo-induced increases to measured PL lifetimes suggest suppression of fast non-radiative pathways rather than purely addition to the total carrier population. Studies involving the quantification of \ch{MoS2}’s PL emission spectra are cautioned to control for light exposure to more accurately compare chemical and electrical treatments on S-TMD optics. 
TFSI treatment has a synergistic effect with photo-enhancement on PL intensity observed in air while both TFSI-treated and untreated samples display no change in vacuum with laser exposure. The results of this work hope to provide insight and support toward understanding inherent mechanisms at play when engineering efficient and effective 2D TMD-based optoelectronics.

\section{Experimental Methods}

\subsection{Preparation of \ch{MoS2} Monolayers}
Bulk \ch{MoS2} crystals are purchased from 2D Semiconductors. Exfoliated \ch{MoS2} monolayers are fabricated using the `gold-assisted exfoliation' method. PL and electrical conductivity of monolayer \ch{MoS2} using this method are equivalent to the `scotch-tape' exfoliation, implying minimal impact on crystal quality \cite{Desai2016}. 100-orientation silicon wafers with 286\,nm of \ch{SiO2} were cleaned with acetone, isopropyl alcohol (IPA), and deionized (DI) water followed by oxygen plasma for $60\,\unit{s}$ in order to create a clean interface. The oxide thickness allows optimal contrast on optical microscopy between single- and multi-layer TMD flakes. Bulk \ch{MoS2} exfoliated by tape had $70\,\unit{\nano m}$ of gold evaporated onto the bulk surface via physical vapour deposition. Thermal release tape was used to lift the gold film and attach monolayer \ch{MoS2} onto a clean silicon wafer. The wafer was heated until the tape was released. Iodine-based (KI) gold etchant (Sigma Aldrich) was used to remove the gold film leaving monolayer \ch{MoS2} on clean silicon. Residual adhesive from the thermal release tape is removed with an additional acetone bath for $1\unit{hr}$ and thermal annealing at $350\unit{^\circ C}$ in argon gas with 5\% hydrogen gas. To remove bulk \ch{MoS2} debris, a polypropylene carbonate (PPC) and polymethylsiloxane (PDMS) film is used in conjunction with a micromanipulator system to pick up bulk material and left-over debris while leaving the thin monolayer \ch{MoS2} behind. The PPC/PDMS film is gentle enough to not detach any monolayer \ch{MoS2} which is strongly adhered to the Si\ch{O2} substrate via thermal annealing. AFM measurements to verify monolayer thickness are measured on a Bruker Icon AFM in non-contact tapping mode. 

\subsection{Chemical Treatments}
All TFSI treatments are performed in a positive-pressure, nitrogen-filled glove box. $50\,\unit{\milli g}$ of TFSI was diluted in  $10\,\unit{\milli g}$ of dichloroethane (DCE) to create a $5.0\,\unit{\milli g/\milli L}$ solution. A further 1:10 dilution of dichlorobenzene (DCB) was used to create a $0.5\,\unit{\milli g/\milli L}$ solution. \ch{MoS2} samples are placed in TFSI solution and sealed in a vial for $1\,\unit{hr}$. Samples are nitrogen blow-dried followed by a $5\,\unit{min}$ annealing step at $80\unit{^\circ C}$ on a hotplate in ambient conditions to dry the remaining wet acid. Optical characterization and measurements are carried out immediately after TFSI treatment in order to limit the degradation of the TFSI coating and its exposure to light. 

\subsection{Photoluminescence Characterization}
CW PL measurements are performed on a home-built microPL system, equipped with a Thorlabs CPS532 diode laser ($532\,\unit{\nano m}$ center wavelength) as an excitation light source. Excitation spot size is $\sim$2.0\,\unit{\micro m} diameter FWHM unless otherwise stated. The objective lens used in all measurements is a Thorlabs C240TMD-B (NA\,0.5, f\,=\,$8.0\,\unit{\milli m}$). Fluorescence is analyzed using a Jobin Yvon iHR320 Imaging Spectrometer equipped with an Andor iXon electrically-cooled charge coupled device (CCD) camera. The system can be switched between a white-light imaging mode for locating samples with sub-$2\,\unit{\micro m}$ accuracy, and the CW PL measurement mode.

SPCL measurements are performed using an ultrafast light source coupled into the same microPL system used for CW PL measurements. We use $1100\,\unit{\nano m}$, $300\,\unit{\femto s}$ duration laser pulses generated by a Light Conversion Pharos/Orpheus Laser and optical parametric amplifier (OPA) system at 1\,MHz rep rate. The $1100\,\unit{\nano m}$ OPA output is doubled to $550\,\unit{\nano m}$ via second harmonic generation (SHG) in an external $1.0\,\unit{\milli m}$-thick anti-reflective (AR)-coated beta barium borate (BBO) crystal, with a high-resolution motorized rotation stage and manual rotation mount for angle tuning in two degrees of freedom. Excitation spot size is $\sim$$2.0\,\unit{\micro m}$ diameter FWHM. Fluorescence from the pulsed excitation is bandpass filtered from $630\,\unit{\nano m}$ to $670\,\unit{\nano m}$ and delivered to a Perkin-Elmer single-photon avalanche photodiode connected to a qutag time-to-digital converter. The IRF is measured by moving to a region of the sample with only bare substrate, removing the longpass filter to permit transmission of the $550\,\unit{\nano m}$ laser, and reduction of a variable attenuator until a signal is detected with similar count rates to the histograms obtained from \ch{MoS2} PL.

\begin{suppinfo}
The Supporting Information is available free of charge.

\begin{itemize}
  \item Supplemental.pdf: sample-wide PL peak wavelength changes due to acid treatment and laser exposure, effects of annealing, additional power-dependent PL measurements, additional SPCL measurements and analysis at independent locations. (PDF)
\end{itemize}

\end{suppinfo}

\section{Author Contributions}
Authors KHT and JRG contributed equally to this effort. The project was designed and managed by RGK (supervising KHT) and JMF (supervising JRG). KHT fabricated all samples, conducted AFM measurements, and performed TFSI treatments. JRG conducted CW PL and ultrafast SPCL characterization with assistance from KHT. Data analysis and visualization was performed by KHT and JRG. Data interpretation was contributed by all authors. The manuscript was drafted by JRG and KHT, and all authors participated in the manuscript revision process.

\begin{acknowledgement}

We acknowledge support from the Natural Sciences and Engineering Research Council of Canada through the Discovery Grants program [$\#$2018-05192] and the Collaborative Research and Training Experience program (CREATE-Materials for Advanced Photonics and Sensing [$\#$2018-511093]), and the Canada Foundation for Innovation [$\#$2017-36423]. Fabrication was possible thanks to NanoFabrication Kingston facilities as well as AFM measurements at the Queen's Surface Facility in the Department of Chemistry. 

\end{acknowledgement}

\bibliography{MainPaper.bib}

\end{document}


\newpage
\section{2D PL Map of Untreated and Treated Samples}

\begin{figure}[ht!]
\centering\includegraphics[width = 16.5cm, trim={0.2cm 0.50cm 0.15cm 0.45cm},clip,keepaspectratio=true]{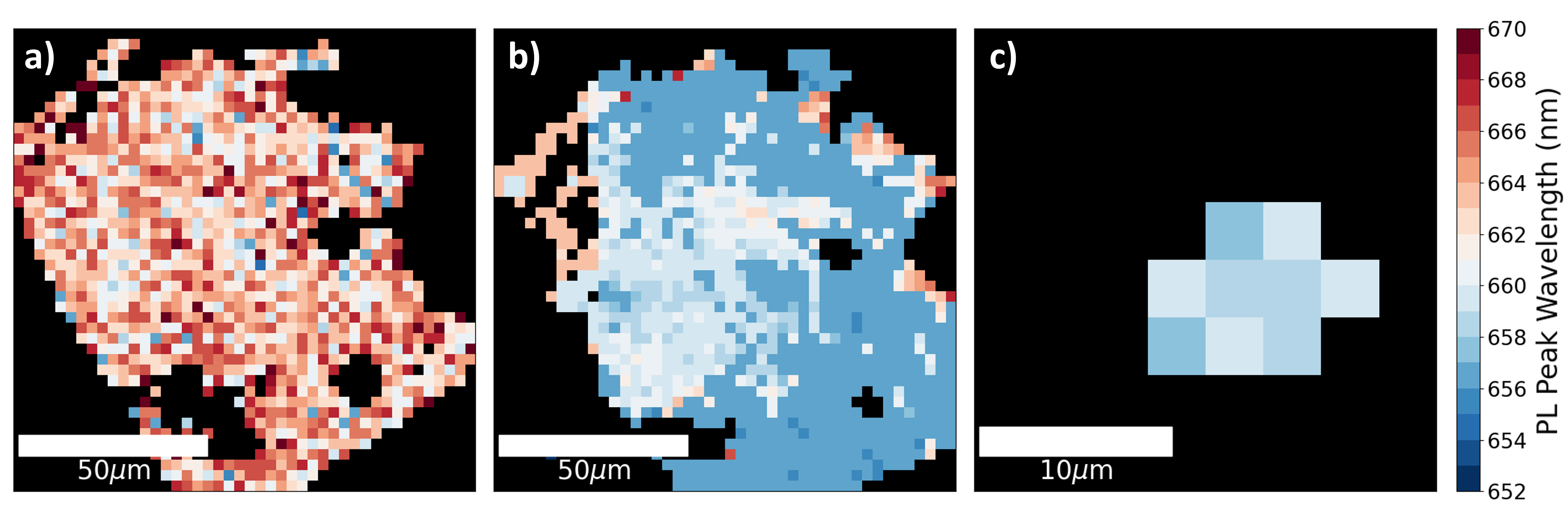}
\caption{PL peak wavelength location 2D raster scan of the sample used in \F1. Peak wavelength of PL in a 2D map of (a) untreated, (b) TFSI-treated MoS$_2$, and (c) TFSI-treated and laser-enhanced MoS$_2$.  Areas with low PL signal are omitted (e.g., substrate and unenhanced portions in (c)). 2D map reveals a semi-uniform blue shifting during acid treatment.}
\end{figure}

\newpage
\section{Stability of TFSI-Treated MoS$_2$ Over Time With No Light Exposure}

PL measurements of a TFSI-treated MoS$_2$ sample were taken at several-hour intervals with 265\,\unit{\micro W} of 532\,nm CW excitation show as \F S2. The laser was turned off between measurements to limit the amount of laser exposure on the sample. The TFSI-enhanced PL intensity does not change significantly over 150\,hr indicating the TFSI acid layer has not yet deteriorated. With minimal laser exposure (<10\,s  per point), the PL intensity remains consistent over time. We found PL photo-enhancement still occurred days after the initial TFSI-bath.  

\begin{figure}[ht!]
\centering\includegraphics[width = 12cm,clip,keepaspectratio=true]{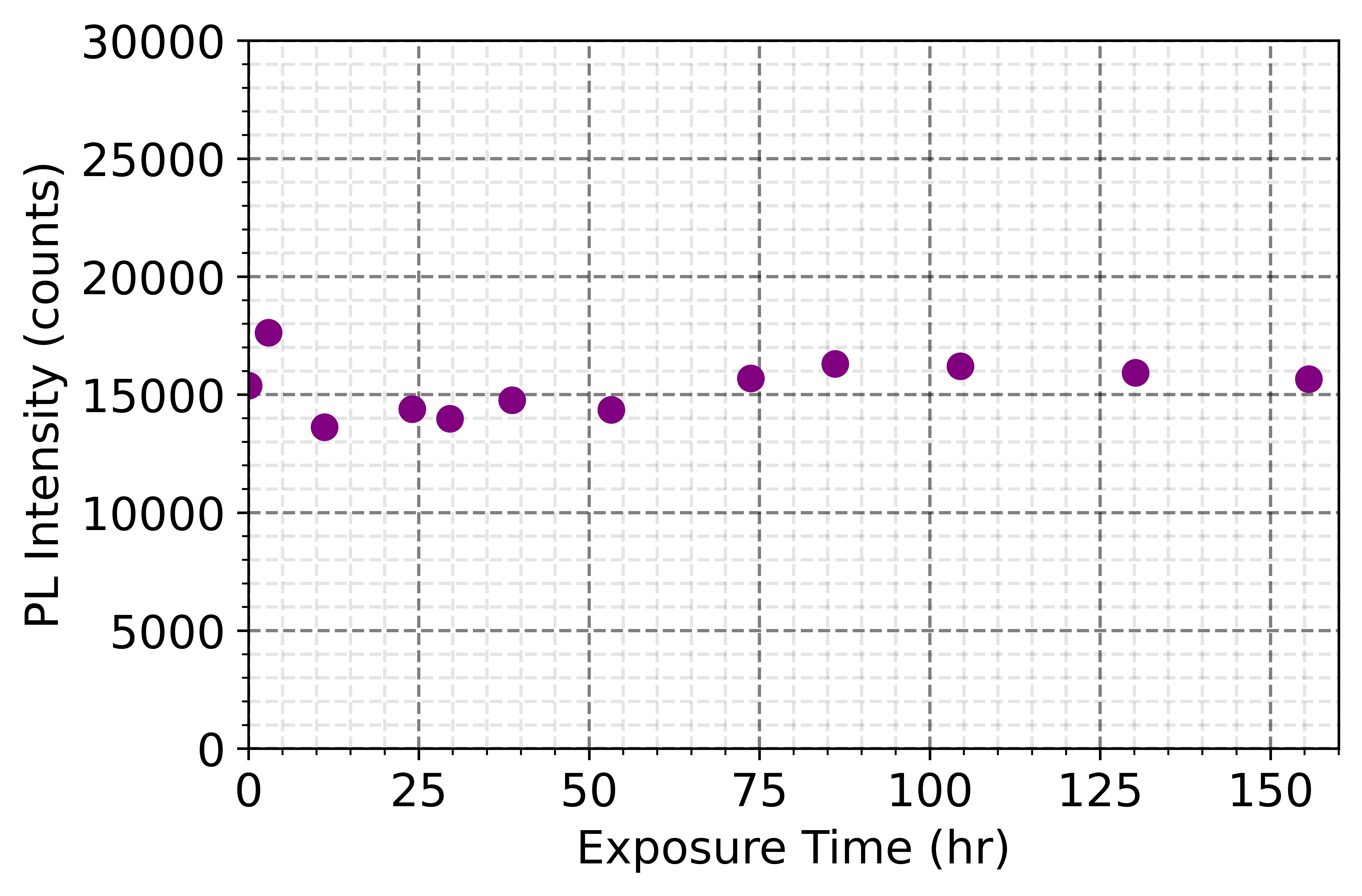}
\caption{PL intensity over time of TFSI-treated MoS$_2$ measured intermittently with no light exposure between measurements. PL intensity reported as the integrated counts of the PL spectrum measured from 615\,nm to 725\,nm.}
\end{figure}

\newpage
\section{Effect of Annealing on Untreated PL}

To ensure PL enhancement is primarily due to photoactivation and not localized heating and annealing due to the laser exposure, we compare the long term PL emission of an untreated \ch{MoS2} sample before and after annealing at $120\unit{^\circ C}$ (\F S3). The sample exhibits similar magnitudes of PL enhancement with 532\,nm CW laser exposure prior to and after annealing. The rate of PL increase of both is within a typical range that we observe between different sites on a single \ch{MoS2} flake. 

\begin{figure}[ht!]
\centering\includegraphics[width = 12cm, trim={0.25cm 0.20cm 0.30cm 0.25cm},clip,keepaspectratio=true]{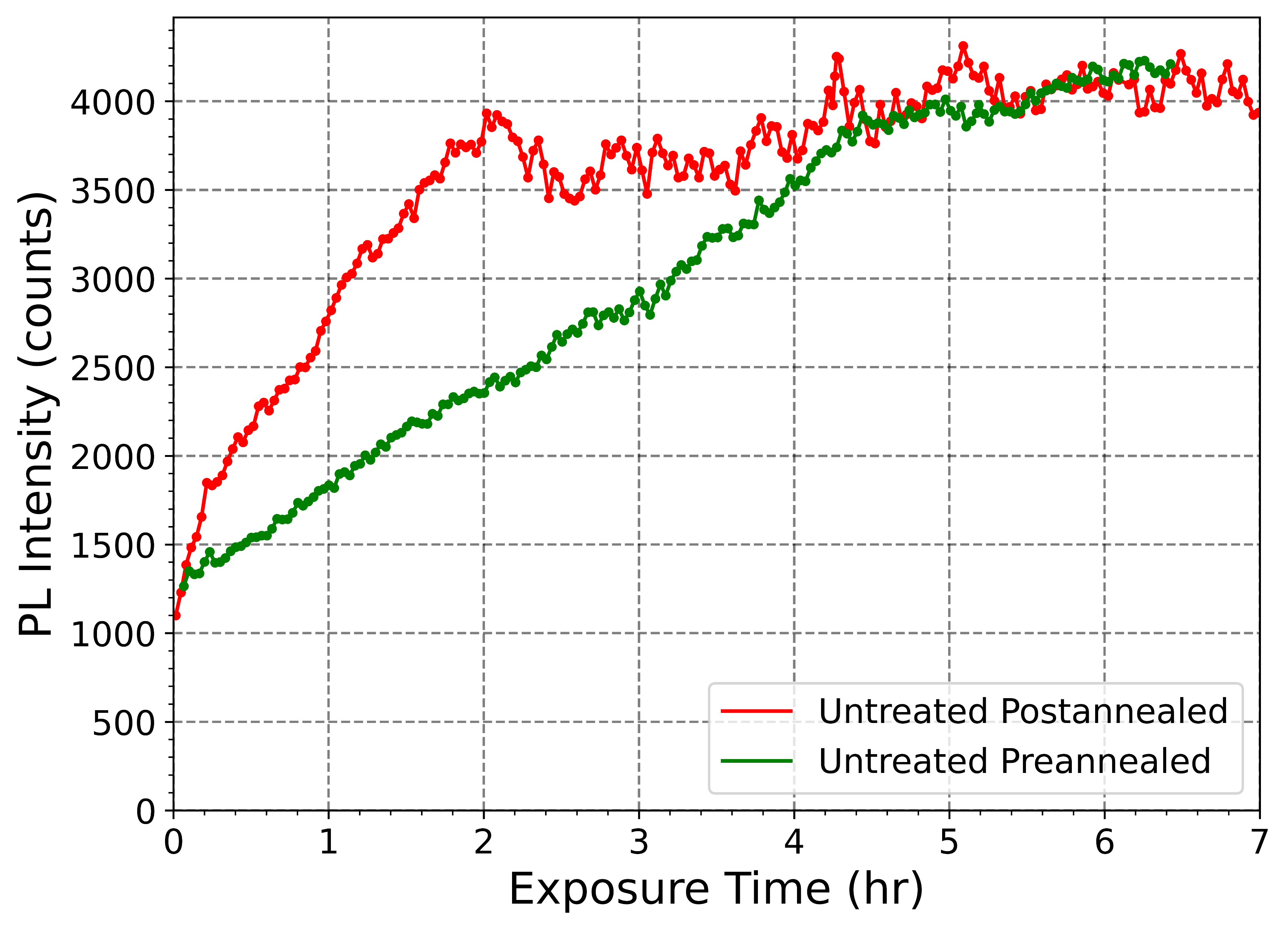}
\caption{PL intensity of untreated MoS$_2$ monolayer under light exposure (CW 532\,nm), without and with annealing at 120$^{\circ}$C in ambient atmosphere. No significant changes to the final PL intensity were observed after 6\,hr of PL tracking.}
\end{figure}

\newpage
\section{Power-Dependent PL of \ch{MoS2} Under CW Excitation}
The power dependence of PL from untreated and TFSI-treated MoS$_2$ was investigated, given as \F S4. With our samples being prone to modification by the measurements - especially the acid-treated sample - care was taken to limit the laser exposure to what was strictly necessary for the measurement. Further, increasing and decreasing power measurements (increasing first) were performed to capture any changes caused by the laser dose.
\begin{figure}[ht!]
\centering\includegraphics[width = 16.5cm, trim={0.2cm 0.50cm 0.15cm 0.45cm},clip,keepaspectratio=true]{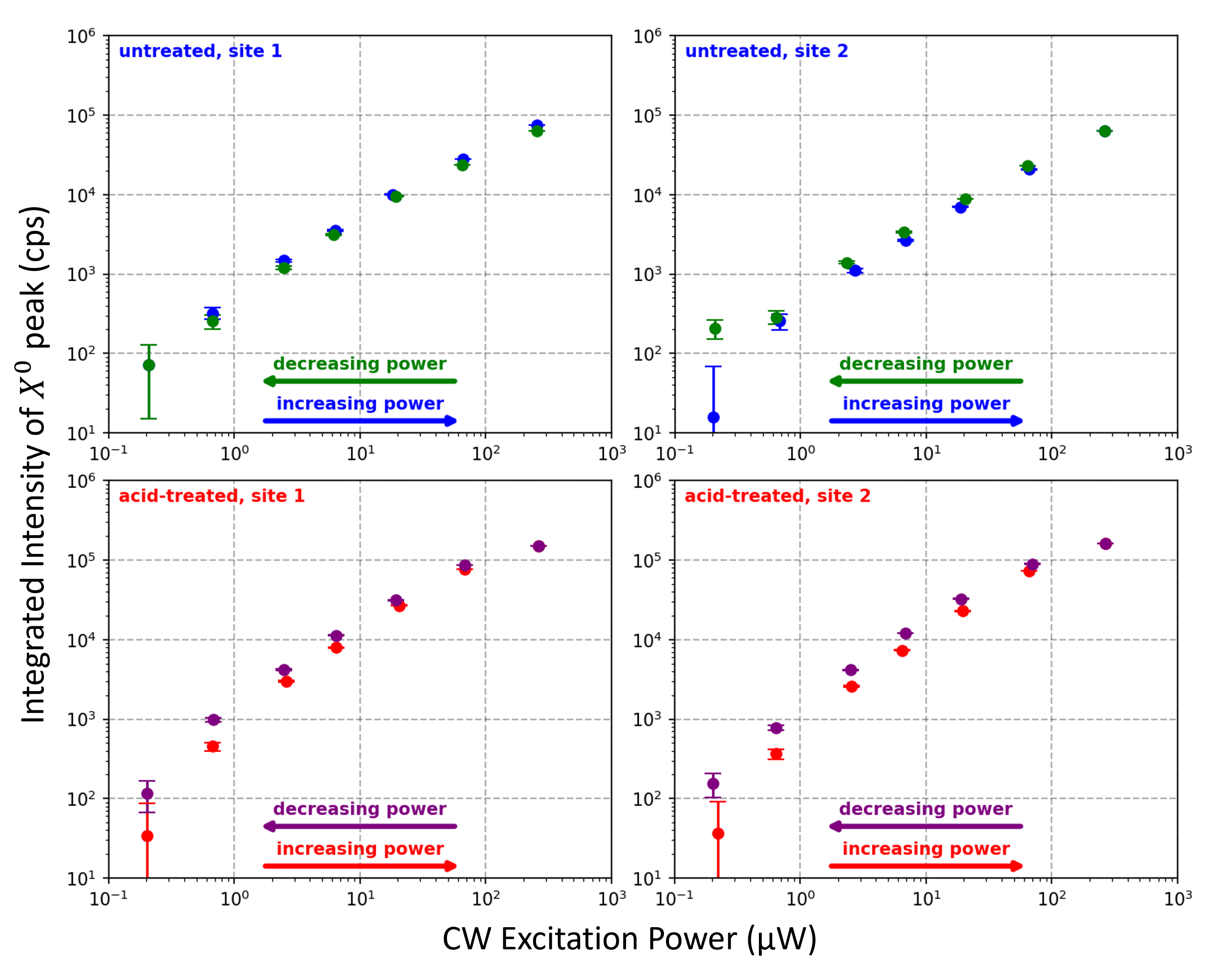}
\caption{Power-dependent PL from the integrated neutral exciton peak ($X^0$) in untreated and TFSI-treated MoS$_2$. All spectra were integrated from 655\,nm to 665\,nm, around the $X^0$ feature at approximately 660\,nm. Acid treated samples with minimal light exposure show less than an order of magnitude enhancement of untreated samples. Most of the explored power range (e.g., <100\,\unit{\micro W}) for treated samples show a linear power dependence.}
\end{figure}

\newpage
\section{Defect Repair and PL Action Cross-Section Growth Rates}

As a way of exploring flux-dependent defect repair, we consider a simple model: defects quench PL for injected excitons within an interaction length of the defect, but the incident flux, over time, repairs the finite number of defects present. Thus, PL intensity scales as 
\begin{equation}
    \label{eqn:PLscaling}
    I_{\mathrm{PL}} \propto I_{\mathrm{exc}} \cdot
    \bigl( K - D\left(t\right)  \bigr),
\end{equation}
where $I_{\mathrm{exc}}$ is the incident excitation intensity, $K$ is the area of the sample being excited, $D\left(t\right)$ is the area within the excitation beam where PL is quenched by defects, and $I_{\mathrm{PL}}$ is the resulting PL intensity. More specifically, the aforementioned PL quenching is likely attributed to trion-induced non-radiative recombination \cite{Amani2015a, Goodman2017}. This model is consistent with the CW measurements reported in the main work that show a linear dependence with excitation over short time periods of laser exposure. Thus, the PL per unit excitation intensity is directly proportional to the bright area of within the excitation spot $K - D\left(t\right)$ which grows over time with incident flux. The PL action cross-section (PLAC) growth rate $R_{\mathrm{PLAC}}$ (the time derivative of the excitation-normalized PL intensity) is proportional to the derivative of the defect area:
\begin{equation}
    \label{eqn:dPLoverIdtscaling}
    R_{\mathrm{PLAC}} = \frac{\mathrm{d}}{\mathrm{d}t}
    \left( \frac{I_{\mathrm{PL}}}{I_{\mathrm{exc}}}\right)
    \propto 
    - \frac{\mathrm{d} D}{\mathrm{d}t}.
\end{equation}

We propose a model where the derivative of the defect-affected area (i.e., the defect repair rate, $R_\mathrm{DR}$) is directly proportional to incident intensity. This is reasonable enough as a simple model for what we assume is a sitewise photomediated repair or compensation process wherein an increased number photons involved means more frequent opportunities to repair or mask a defect. Thus the defect repair rate $R_\mathrm{DR}$ can be defined as:
\begin{equation}
    \label{eqn:DRR}
    R_{\mathrm{DR}} \equiv - \frac{\mathrm{d}D}{\mathrm{d}t}
    = A \cdot D\left( t \right) \cdot I_{\mathrm{exc}},
\end{equation}
where $A$ is a scaling factor that describes the rate at which an incident photon is absorbed and results in the repair of a defect.

Solving the first-order ODE for $D(t)$:
\begin{equation}
\begin{aligned}[b]
    \label{eqn:DofT}
    D(t) &= D(0)\cdot e^{-AI_{\mathrm{exc}}t}.
\end{aligned}
\end{equation}
The result is a defect affected area that decays exponentially -  the repair is the fastest at the onset, when the chance for a photon to interact (via an intermediary, such as an injected exciton) with a defect site is the highest. We substitute back into Eq.\ref{eqn:dPLoverIdtscaling}  and take the natural log to linearize the PLAC rate in time:

\begin{equation}
\begin{aligned}
    \label{eqn:PLAClinearized}
    R_{\mathrm{PLAC}} &= -D(0) A I_{\mathrm{exc}}
    \cdot e^{-A I_{\mathrm{exc}} t}, \\
    \ln{\left( R_{\mathrm{PLAC}} \right)} &= \ln{\left( -D(0) A I_{\mathrm{exc}} \right)} -AI_{\mathrm{exc}} \cdot t.
\end{aligned}
\end{equation}
Thus we can take the natural log of our experimentally-obtained PLAC rate, and from the linear regression on the first 2.5\,hrs of data (\F3a(ii)) we can obtain the $A$ constant in our defect repair model, for which the middle and highest power sets (25.41\,\unit{\micro W} and 118.74\,\unit{\micro W}) agree within 15 percent. However, as the slope in the model is also proportional to excitation intensity, the expected signal for the lowest power set is 10$\times$ weaker than the next, and hence the data suffers from poor signal-to-noise.

Since the PLAC rates are obtained from numerical derivatives, we employ a combination of binning, rolling average smoothing and LOESS filter smoothing to mitigate the effects of measurement noise on the derivative. Particularly, before taking a numerical derivative, all PL data (reported as the mean of the top 100 counts in each spectrum) in \F3a(i) is first smoothed by a LOESS filter (20-pt window, 2/3 of points used).  The first derivative of this data is taken via linear regression fitting on rolling windows of the smoothed PL data (100-pt window for 2.49\,\unit{\micro W}, 10-pt window for 25.41\,\unit{\micro W} and 118.74\,\unit{\micro W}). The resulting derivatives are binned in windows of 20 (80) for the 25.41\,\unit{\micro W} and 118.74\,\unit{\micro W} sets (2.49\,\unit{\micro W} set) to suppress oversampling of the smoothed PL and first derivative data. Ultimately the 2.49\,\unit{\micro W} dataset is far too low signal-to-noise to achieve a particularly meaningful measure of the PLAC rate through a discrete first derivative, as is already indicated through the considerably more aggressive smoothing required to even obtain any reasonable first derivative data.

Similar long-term PL time series of multiple independent neighbouring points averaged together would likely allow one to obtain a measure of the PLAC rate under these conditions with similar signal-to-noise as the 25.41\,\unit{\micro W} and 118.74\,\unit{\micro W} sets.

\newpage
\section{SPCL Measurements on TFSI-Treated Samples}
The same dataset shown in \F 4b is presented alternatively as \F S5, showing correlations of extracted long lifetime and intensity vs. experiment runtime in the leftmost panels. The correlation of extracted lifetime with intensity is shown as the rightmost panel. 
Similar to \F 4b, the extracted SPCL lifetime vs runtime and vs. intensity for 3 other independent sites on the same sample are shown as \F S6. The middle and right panels show that at lower intensities - when we expect the lifetime is also lower based on our repeated observations of increasing detected lifetime with PL intensity - the fitting algorithm results in large scatter between subsequent measurements. Above an intensity of 25\,kcps the fitting obtains results with minimal scatter.
\begin{figure}[ht!]
\centering\includegraphics[width = 14.5cm, trim={0.25cm 0.0cm 0.25cm 0.0cm},clip,keepaspectratio=true]{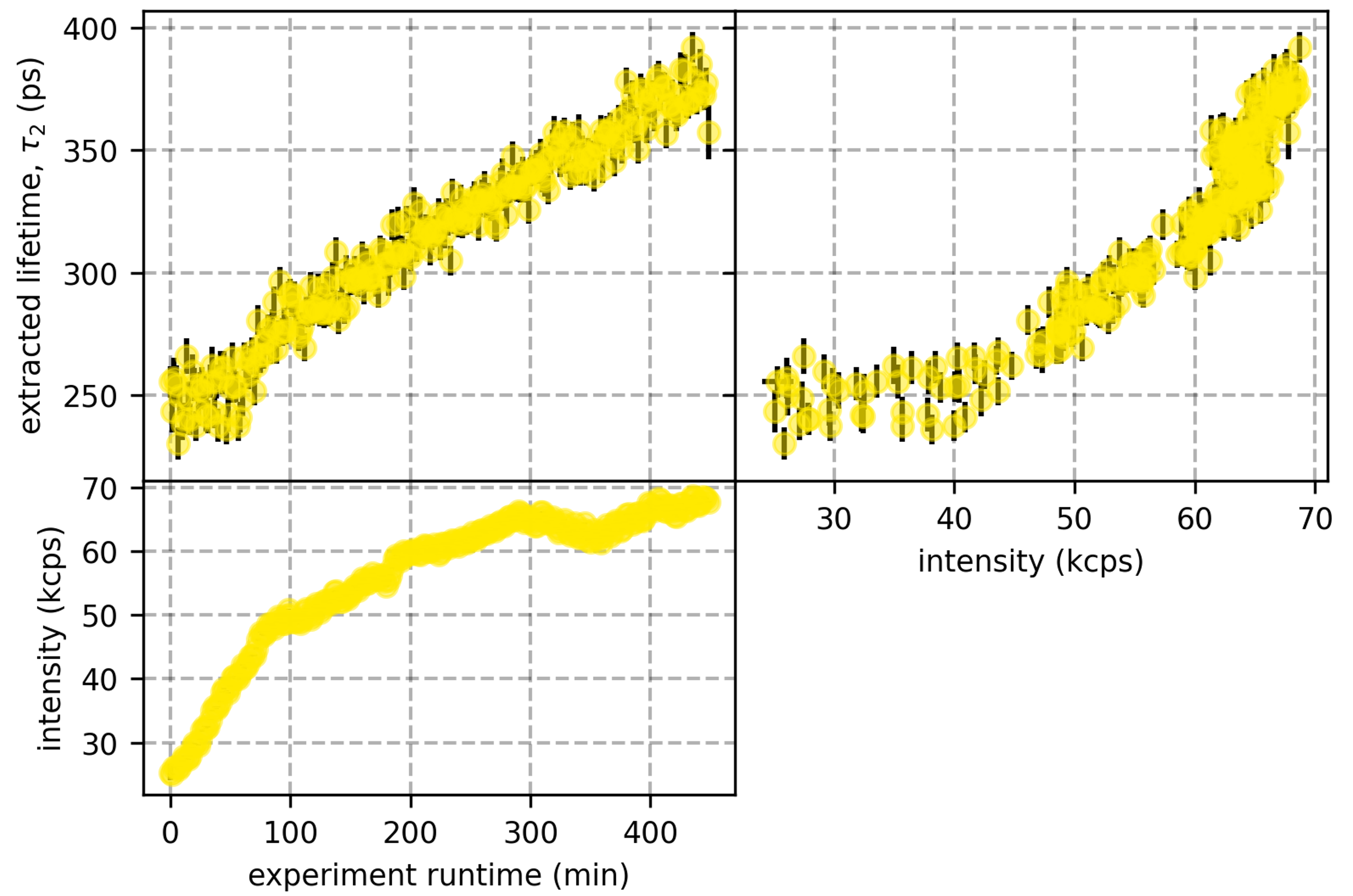}
\caption{A display of different metrics obtained via the SPCL measurement. This is the same single point measurement shown in \F 4b, with the intensity vs. runtime shown explicitly in the bottom left panel. }
\end{figure}
\begin{figure}[ht!]
\centering\includegraphics[width = 16.5cm, trim={0.25cm 0.15cm 0.15cm 0.35cm},clip, keepaspectratio=true]{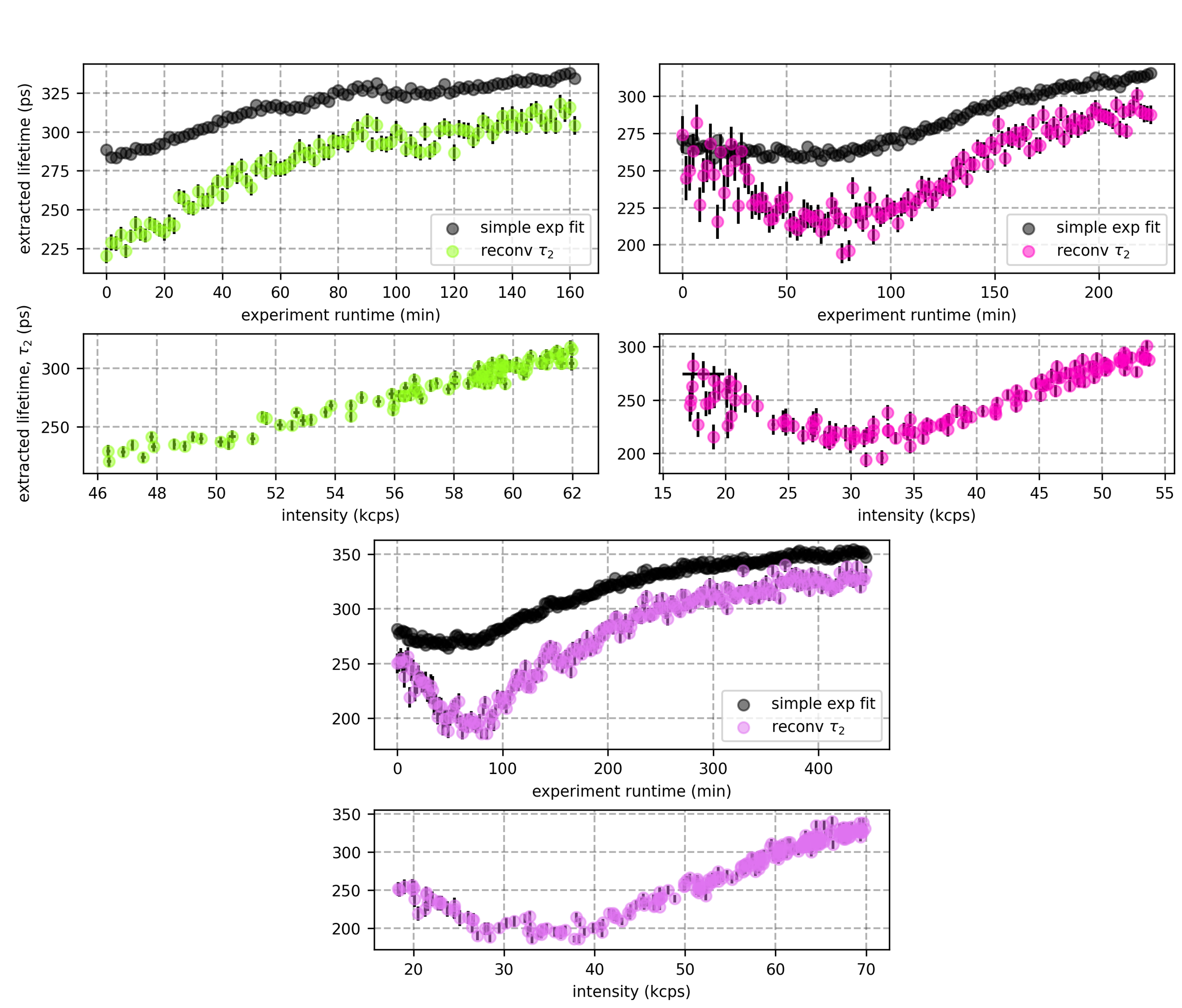}
\caption{SPCL measurements taken on 3 independent locations of the TFSI-treated \ch{MoS2} sample, using both exponential tail-fitting and biexponential reconvolution methods.}
\end{figure}
The fitting results for the multi-parameter function given as Equation (1) as given as \F S7, showing 4 different sites on sample B0, with the colour coding for the different sites matching those of \F S5 and S6. The fitting in all cases shows a modest chance to the extracted short lifetime ranging from 20-30\,ps at the start to approximately 40\,ps, while the extracted long lifetime shows an increase from 250\,ps or below to approximately 300\,ps-375\,ps. A mono-exponential fit function achieved poor results fitting this data.
\begin{figure}[ht!]

\centering\includegraphics[width = 15cm,trim={0.15cm 0.0cm 0.15cm 0.35cm},clip,keepaspectratio=true]{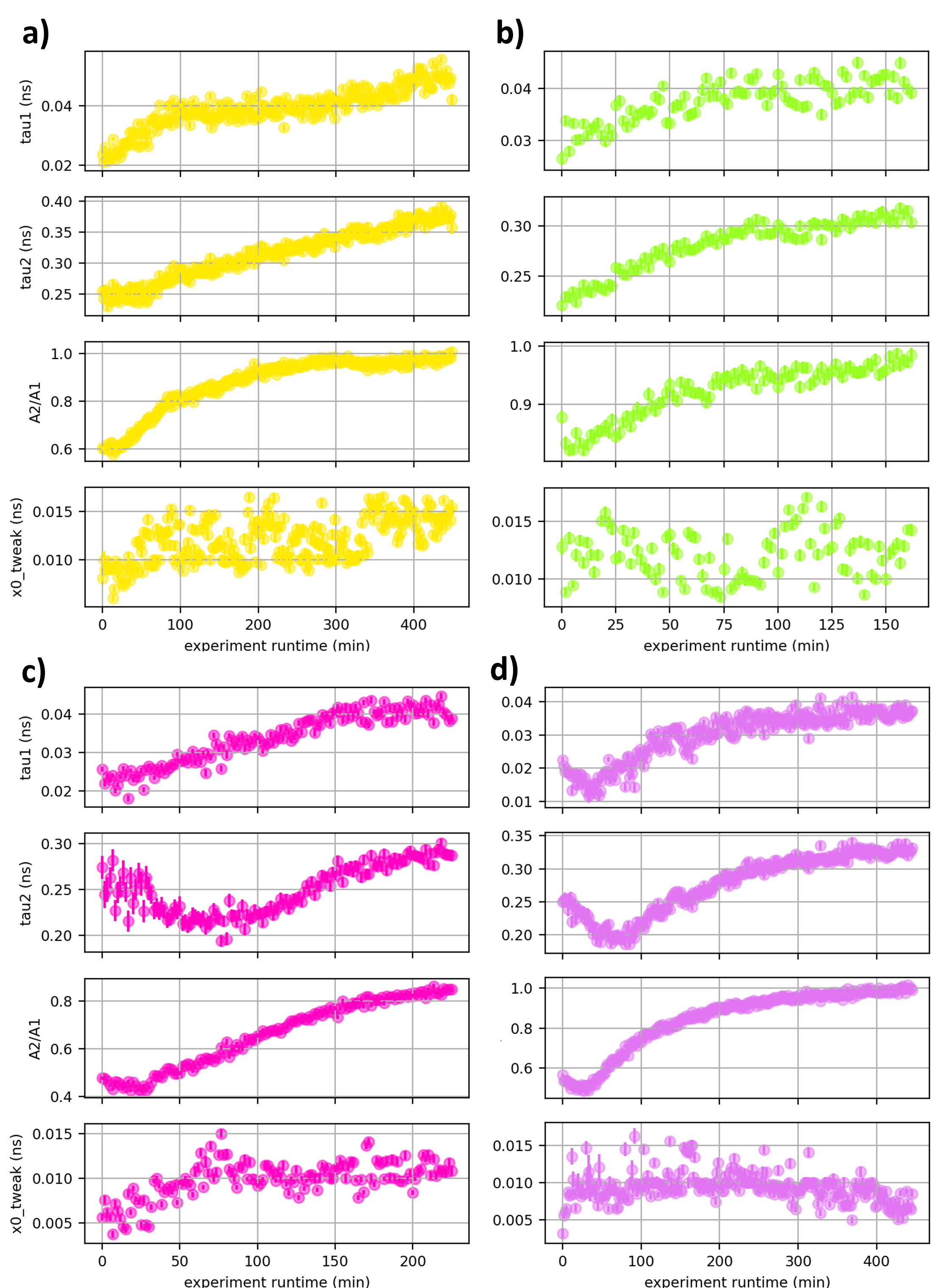}
\caption{Fitting results for the 4-parameter biexponential reconvolution fit for TFSI-treated MoS$_2$. a) shows the fit results for the dataset in \F 4, while b)-d) show the fits for the other 3 locations in \F S6. Note that the x0\_tweak parameter corresponds to $t_0$ in Eq. (1).}
\end{figure}

\clearpage

\bibliography{MainPaper.bib}